\documentclass[prl,aps,preprintnumbers,nofootinbib,twocolumn]{revtex4}
\usepackage{epsfig}
\usepackage{amsmath}
\usepackage{hyperref}

\def\as{\alpha_S}
\def\Lbeta{{\rm L}_\beta}
\def\Lrho{{\rm L}_\rho}
\def\b{\beta}

\begin{document}

\preprint{CERN-PH-TH/2013-056,  TTK-13-08}

\renewcommand{\thefigure}{\arabic{figure}}

\title{The total top quark pair production cross-section at hadron colliders through ${\cal O}(\alpha_S^4)$}

\author{Micha\l{}  Czakon}
\author{Paul Fiedler}
\affiliation{Institut f\"ur Theoretische Teilchenphysik und Kosmologie,
RWTH Aachen University, D-52056 Aachen, Germany}

\author{Alexander Mitov}
\affiliation{Theory Division, CERN, CH-1211 Geneva 23, Switzerland}

\date{\today}

\begin{abstract}
We compute the next-to-next-to-leading order (NNLO) QCD correction to the total cross-section for the reaction $gg \to t\bar t + X$. Together with the partonic channels we computed previously, the result derived in this letter completes the set of NNLO QCD corrections to the total top pair production cross-section at hadron colliders. Supplementing the fixed order results with soft-gluon resummation with next-to-next-to-leading logarithmic accuracy we estimate that the theoretical uncertainty of this observable due to unknown higher order corrections is about 3\% at the LHC and 2.2\% at the Tevatron. We observe a good agreement between the Standard Model predictions and the available experimental measurements. The very high theoretical precision of this observable allows a new level of scrutiny in parton distribution functions and new physics searches.  
\end{abstract}
\maketitle

\section{Introduction}

Production of top quark pairs at hadron colliders is among the processes that are most challenging to theory. Bringing this process under good theoretical control therefore represents a significant step in our broader understanding of perturbative Quantum Chromodynamics (QCD) and its applications at hadron colliders. 

The first step in this direction was made some 25 years ago, when the next-to-leading order (NLO) QCD corrections to $t\bar t$ production were computed in the groundbreaking works \cite{Nason:1987xz,Beenakker:1988bq}. The complexity of the NLO calculations required the application of purely numerical methods, and it took almost twenty years before the exact analytic result appeared \cite{Czakon:2008ii} revealing the full complexity of the cross-section for massive fermion hadroproduction. 

In the last few years we are witnessing a significant interest in computing next-to-next-to leading order (NNLO) corrections to hadron collider processes. Such a demand is dictated in part by the high-precision measurements available from the LHC and the Tevatron. The first hadron collider processes that were computed at NNLO, namely, Drell-Yan and vector boson \cite{Hamberg:1990np,Anastasiou:2003yy,Anastasiou:2003ds}, Higgs \cite{Harlander:2002wh,Anastasiou:2002yz,Ravindran:2003um} and diphoton \cite{Catani:2011qz} production, all share the properties of (a) having massless QCD partons and (b) involving at leading order (LO) two partons meeting in a color singlet vertex. Tackling processes with higher complexity, among which $t\bar t$ production is a prominent example, proved to require new computational approaches.

About one year ago, the first step in this direction was made precisely in the context of $t\bar t$ production. Based on a new view \cite{Czakon:2010td} about how to treat double-real radiation corrections, the first genuinely NNLO corrections to the total inclusive cross-section in $q\bar q\to t\bar t+X$ were computed \cite{Baernreuther:2012ws}. Later on, the partonic reactions involving at least one fermion in the initial state were also completed \cite{Czakon:2012zr,Czakon:2012pz}. In this work we report the calculation of the last missing NNLO correction to $t\bar t$ production, in the partonic reaction $gg\to t\bar t+X$. With this calculation, the complete set of NNLO corrections to the total inclusive cross-section for top pair production at hadron colliders is now known. In this letter, for the first time, we quantify their phenomenological implications. 

Before closing this section we would like to point out the very recent NNLO calculation of the process $pp\to H+j$ \cite{Boughezal:2013uia} which was performed with methods similar to ours and, in particular, the subtraction scheme proposed by one of us \cite{Czakon:2010td}. Moreover, a first partial result for dijet production $pp\to jj$ at NNLO has just appeared \cite{Ridder:2013mf}. We believe that this burst of precision applications at hadron colliders marks the outset of a new and lasting stage in precision physics at hadron colliders.

\section{The $t\bar t$ production cross-section}

In this letter we consider the total inclusive $t\bar t$ production cross-section
\begin{equation}
\sigma_{\rm tot} = \sum_{i,j} \int_0^{\beta_{\rm max}}d\beta\, \Phi_{ij}(\beta,\mu_F^2)\, \hat\sigma_{ij}(\beta,m^2,\mu_F^2,\mu_R^2) \, .
\label{eq:sigmatot}
\end{equation}
The indices $i,j$ run over all possible initial state partons; $\beta_{\rm max}\equiv\sqrt{1-4m^2/S}$; $\sqrt{S}$ is the c.m. energy of the hadron collider and $\beta=\sqrt{1-\rho}$, with $\rho\equiv 4m^2/s$, is the relative velocity of the final state top quarks with pole mass $m$ and partonic c.m. energy $\sqrt{s}$.

The function $\Phi$ in Eq.~(\ref{eq:sigmatot}) is the partonic flux 
\begin{equation}
\Phi_{ij}(\beta,\mu_F^2) = {2\beta \over 1-\beta^2}~ {\cal L}_{ij}\left({1-\beta_{\rm max}^2\over 1-\beta^2}, \mu_F^2\right) \, ,
\label{eq:flux}
\end{equation}
expressed through the usual partonic luminosity
\begin{eqnarray}
{\cal L}_{ij}(x,\mu_F^2) = x \left( f_i\otimes f_j \right) (x,\mu_F^2) \, .
\label{eq:Luminosity}
\end{eqnarray}

As usual, $\mu_{R,F}$ are the renormalization and factorization scales. Setting $\mu_F=\mu_R=m$, the NNLO partonic cross-section can be expanded through NNLO as
\begin{eqnarray}
\hat\sigma_{ij}\left(\beta\right) = {\as^2\over m^2}\left(  \sigma^{(0)}_{ij} + \as\sigma^{(1)}_{ij} + \as^2 \sigma^{(2)}_{ij} + {\cal O}(\as^3)\right) \, .
\label{eq:sigmapart}
\end{eqnarray}
In the above equation $\as$ is the ${\overline{\rm MS}}$ coupling renormalized with $N_L=5$ active flavors at scale $\mu_R^2=m^2$ and $\sigma^{(n)}_{ij}$ are functions only of $\beta$. The procedure for restoring the dependence on $\mu_F\neq\mu_R\neq m$ is standard and has been detailed, for example, in Ref.~\cite{Czakon:2012pz}.

All partonic cross-sections are known exactly through NLO \cite{Nason:1987xz,Beenakker:1988bq,Czakon:2008ii}. The NNLO corrections to the partonic reactions $ij=q\bar q,\,qg,\,qq,\,qq',\,q\bar q'$ were computed in Refs.~\cite{Baernreuther:2012ws,Czakon:2012zr,Czakon:2012pz}. In the following we present the results for $ij=gg$.

\section{Parton level results for  $gg\to t\bar t +X$}

Keeping the dependence on the number of light flavors $N_L$ explicit, the NNLO correction $\sigma^{(2)}_{gg}$ reads
\begin{eqnarray}
\sigma^{(2)}_{gg}(\beta) = F_0(\beta) +F_1(\beta) N_L +F_2(\beta) N_L^2  \, .
\label{eq:sigma2gg}
\end{eqnarray}
The functions $F_i\equiv F^{(\beta)}_i+F^{(\rm fit)}_i,~i=0,1,2,$ read:
\begin{eqnarray}
F^{(\beta)}_2 &=& 0 \, , \label{eq:F2beta0}\\ 
F^{(\beta)}_1 &=& \sigma^{(0)}_{gg}\left[
(-0.00611924+0.0436508 \Lbeta)/\b \right. \nonumber\\
&& \left. +0.139124 \Lbeta-0.755826 \Lbeta^2+0.54038 \Lbeta^3
\right] \, , \label{eq:F1beta0} \\ 
F^{(\beta)}_0 &=&  \sigma^{(0)}_{gg}\left[
0.43408/\b^2+14.8618 \Lbeta-1.99838 \Lbeta^2 \right. \nonumber\\
&&  -14.7016 \Lbeta^3+29.1805 \Lbeta^4 \nonumber\\
&& \left.+(-0.0240072+1.81537 \Lbeta+3.14286 \Lbeta^2)/\b
\right]  \, , \label{eq:F0beta0}\\ 
F^{(\rm fit)}_2 &=& 
10^{-4}\left[(6.44022 \b-4.8664 \b^2-0.0324653 \Lrho^2) \rho \right. \nonumber\\
&&  +(-13.8424 \b+4.7366 \b^2-2.91398 \Lrho) \rho^2 \nonumber\\
&& \left.+(8.43828 \b-2.78748 \b^2+2.38971 \b^3) \rho^3\right]
\, , \label{eq:F2fit}\\ 
F^{(\rm fit)}_1 &=& 
-0.0195046 \b-1.4717 \b^2-0.223616 \b^3 \nonumber\\
&& +0.499196 \b^5+1.32756 \b^7+0.00466872 \b^3 \Lbeta \nonumber\\
&& +0.0321469 \b^6 \Lbeta^2+(0.579781 \Lrho^2+0.166646 \Lrho^3) \rho \nonumber\\
&& +(-1.36644 \Lrho+2.24909 \Lrho^2) \rho^2
\, , \label{eq:F1fit}\\ 
F^{(\rm fit)}_0 &=& 
581.27542 \b + 1251.4057 \b^2 - 60.478096 \b^3 \label{eq:F0fit}\\
&& + 1101.2272 \b^4 -2905.3858 \b^5 + 629.9128 \b^4\Lbeta \nonumber\\ 
&&- 5.1891075 \Lrho  + (1200.741 \Lrho + 162.50333 \Lrho^2) \rho \nonumber\\ 
&& + (36.074524 \Lrho - 1192.8918 \Lrho^2-1810.2849 \b) \rho^2 \nonumber\\ 
&& + 1568.7591 \b \rho^3 - 461.21326 \b \rho^4 + 121.6379 \b \rho^5 \, , \nonumber
\end{eqnarray}
where $\Lrho\equiv\ln(\rho)$ and $\Lbeta\equiv\ln(\beta)$. The functions $F^{(\beta)}_{2,1,0}$ constitute the analytically known threshold approximation to $\sigma^{(2)}_{gg}$ \cite{Beneke:2009ye}, including the exact Born term
\begin{eqnarray}
\sigma^{(0)}_{gg} = {\pi \b \rho \over 192} \left({16+16 \rho+\rho^2\over \b} \ln\left({1+\b\over 1-\b}\right) -28-31 \rho\right),~
\label{eq:born}
\end{eqnarray}
and with the constant $C^{(2)}_{gg}=0$ (as defined in Ref.~\cite{Beneke:2009ye}). 

The functions $F_{2,1,0}$ are computed numerically, in 80 points on the interval $\beta\in (0,1)$. Details about the calculation are given in the next section. 

\begin{figure}[t]
\centering
\hspace{0mm} 
\includegraphics[width=0.49\textwidth]{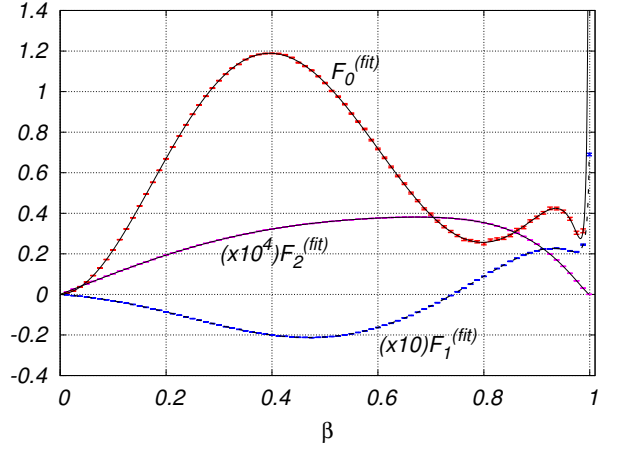} 
\caption{The functions $F^{(\rm fit)}_{2,1,0}$ (\ref{eq:F2fit},\ref{eq:F1fit},\ref{eq:F0fit}) versus the 80 computed points $F_i-F^{(\beta)}_i, i=2,1,0$ (numerical errors at each point are also shown). For improved visibility, the function $F^{(\rm fit)}_1$ is multiplied by a factor of 10, while $F^{(\rm fit)}_2$ by a factor of $10^4$.}
\label{fig:F210-fit}
\end{figure}

Following the approach of Refs.~\cite{Baernreuther:2012ws,Czakon:2012zr,Czakon:2012pz}, the functions $F^{(\rm fit)}_{2,1,0}$ are derived as fits to the difference $F_i-F^{(\beta)}_{2,1,0}$. The functions $F^{(\rm fit)}_i$ together with the discrete values for $F_i-F^{(\beta)}_i$ (including the numerical errors) are shown in fig.~\ref{fig:F210-fit}.
As can be seen from fig.~\ref{fig:F210-fit} the functions $F^{(\rm fit)}_i$ vanish smoothly at threshold $\beta\to 0$, which implies that our calculation agrees with the exactly known threshold behavior \cite{Beneke:2009ye}. This is a strong check of our result.

\begin{figure}[t]
\centering
\hspace{0mm} 
\includegraphics[width=0.49\textwidth]{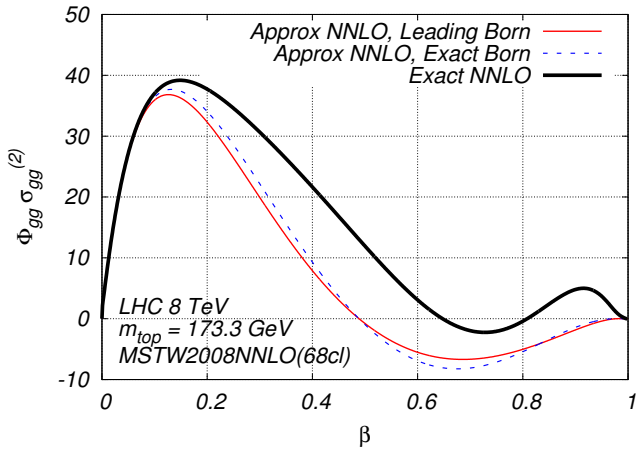} 
\caption{Partonic cross-section times $gg$ flux (\ref{eq:flux}) for the following three cases: exact NNLO (thick black line), approximate NNLO with exact Born term (blue dashed line) and approximate NNLO with leading Born term (thin red line).}
\label{fig:lhc8-fluxes}
\end{figure}

To assess the size of the newly derived NNLO correction, in fig.~\ref{fig:lhc8-fluxes} we compare: (a) the exact NNLO result, (b) the approximate NNLO result with exact Born term and (c) the approximate NNLO result with Born term restricted to its leading power of $\beta$. Each of these three partonic cross-sections is multiplied by the $gg$ partonic flux Eq.~(\ref{eq:flux}) for LHC 8 TeV.  We observe that the power corrections derived in the present work are very large. In fact their contribution to the integrated cross-section is virtually as large as the one due to pure soft gluon corrections. 

The partonic cross-section's leading power behavior in the high-energy limit $\beta\to 1$ reads \cite{Nason:1987xz,Catani:1990xk,Collins:1991ty,Catani:1990eg,Catani:1993ww,Catani:1994sq}
\begin{equation}
\sigma^{(2)}_{gg} \Big\vert_{\rho \to 0} \approx c_1\ln(\rho) + c_0 +{\cal O}(\rho) \, .
\label{eq:high-ebergy-limit}
\end{equation}
The constant $c_1\approx -5.1891075\dots$ is known exactly \cite{Ball:2001pq}. To improve the accuracy of the partonic result (\ref{eq:sigma2gg}) in the high-energy limit, we have imposed on it the logarithmic behavior $\sim c_1\ln(\rho)$ implied by Eq.~(\ref{eq:high-ebergy-limit}). Numerical prediction for the constant term $c_0$ was given in Ref.~\cite{Moch:2012mk}. Our fits return the value $c_0=-31.96+0.1119N_L$ which falls within the range estimated in Ref.~\cite{Moch:2012mk}.

The parton level results derived in this section can be used to derive an estimate for the so-far unknown constant $C^{(2)}_{gg}$ appearing in the threshold approximation \cite{Beneke:2009ye}. Expanding Eq.~\ref{eq:sigma2gg} around the limit $\beta\to 0$ we obtain
\begin{eqnarray}
C^{(2)}_{gg} = 338.179 - 26.8912 N_L + 0.142848 N_L^2 \, .
\label{eq:C2}
\end{eqnarray}

As explained in Ref.~\cite{Hagiwara:2008df}, the estimate (\ref{eq:C2}) for $C^{(2)}_{gg}$ has to be used with caution and a sizable uncertainty should be assumed. We have no good way of estimating the error on the extracted constant and to be reasonably conservative in the following we take this error to be $50\%$. 

The constant $C^{(2)}_{gg}$ is related \cite{Cacciari:2011hy} to the hard matching coefficients $H^{(2)}_{gg,{\bf 1,8}}$ needed for NNLL soft gluon resummation matched to NNLO. However, since our calculation deals with the color averaged cross-section, we cannot extract both constants $H^{(2)}_{gg,{\bf 1,8}}$. We proceed as follows.

Close to threshold, the color singlet and color octet contributions to $\sigma^{(2)}_{gg}$ have independent constant terms $C^{(2)}_{gg,{\bf 1,8}}$, with the constant $C^{(2)}_{gg}$ in Eq.~(\ref{eq:C2}) being their color average. We parameterize the second, unknown, combination of $C^{(2)}_{gg,{\bf 1,8}}$ by their ratio $R^{(2)}_{gg}\equiv C^{(2)}_{gg,{\bf 8}}/C^{(2)}_{gg,{\bf 1}}$, which has the advantage of being normalization independent. For any guessed value of $R^{(2)}_{gg}$, together with Eq.~(\ref{eq:C2}), we can extract values for the hard matching constants $H^{(2)}_{gg,{\bf 1,8}}$. As a guide for a reasonable value of $R^{(2)}_{gg}$ we take the one-loop result (see \cite{Beneke:2009ye,Hagiwara:2008df}): $R^{(1)}_{gg}\equiv C^{(1)}_{gg,{\bf 8}}/C^{(1)}_{gg,{\bf 1}}=2.18$.

In the following we vary $R^{(2)}_{gg}$ in the range $0.1\leq R^{(2)}_{gg}\leq 8$; for each value of $R^{(2)}_{gg}$ we then vary the color averaged constant $C^{(2)}_{gg}$ by additional $50\%$. We observe that as a result of this rather conservative variation, the NNLO+NNLL theoretical prediction for LHC 8 TeV changes by 0.4\% (in central value) and by 0.2\% (in scale dependence). Given the negligible phenomenological impact of these variations, we choose as our default values:
\begin{eqnarray}
H^{(2)}_{gg,{\bf 1}} = 53.17,~~H^{(2)}_{gg,{\bf 8}} = 96.34~~ ({\rm for}\, N_L=5) \, ,
\label{eq:C2H2}
\end{eqnarray}
derived from Eq.~(\ref{eq:C2}) and the mid-range value $R^{(2)}_{gg}=1$.

\section{Calculation of $gg\to t\bar t +X$ through ${\cal O}(\as^4)$}\label{sec:calc}

The calculation of the ${\cal O}(\as^4)$ corrections to $gg\to t\bar t+X$ is performed in complete analogy to the calculations of the remaining partonic reactions \cite{Baernreuther:2012ws,Czakon:2012zr,Czakon:2012pz}. The two-loop virtual corrections are computed in \cite{gg-two-loop}, utilizing the analytical form for the poles \cite{Ferroglia:2009ii}. We have computed the one-loop squared amplitude; it has previously been computed in \cite{Anastasiou:2008vd}. The real-virtual corrections are derived by integrating the one-loop amplitude with a counter-term that regulates it in all singular limits \cite{Bern:1999ry}. The finite part of the one-loop amplitude is computed with a code used in the calculation of $pp\to t\bar t+{\rm jet}$ at NLO \cite{Dittmaier:2007wz}. The double real corrections are computed in \cite{Czakon:2010td}. Factorization of initial state collinear singularities as well as $\mu_{F,R}$ scale dependence is computed in a standard way; see Refs.~\cite{Czakon:2012zr,Czakon:2012pz}.

\section{Phenomenological applications}

In table~\ref{tab:results} we present our most precise predictions for the Tevatron and LHC at 7, 8 and 14 TeV. All numbers are computed for $m=173.3$ GeV and MSTW2008nnlo68cl pdf set~\cite{Martin:2009iq} with the program {\tt Top++}~({\tt v2.0})~\cite{Czakon:2011xx}. Scale uncertainty is determined through independent restricted variation of $\mu_F$ and $\mu_R$.
\begin{table}[h]
\begin{center}
\begin{tabular}{|c|c|c|c|}
\hline
Collider & $\sigma_{\rm tot}$ [pb]  & scales [pb] & pdf [pb] \\
\hline
Tevatron & $7.164$ & $ ^{+0.110 (1.5\%)}_{-0.200 (2.8\%)} $ & $ ^{+0.169 (2.4\%)}_{-0.122 (1.7\%)} $   \\
\hline
LHC 7 TeV & $172.0$ & $ ^{+4.4 (2.6\%)}_{-5.8 (3.4\%)} $ & $ ^{+4.7 (2.7\%)}_{-4.8 (2.8\%)} $  \\
\hline
LHC 8 TeV & $245.8$ & $ ^{+6.2 (2.5\%)}_{-8.4 (3.4\%)} $ & $ ^{+6.2 (2.5\%)}_{-6.4 (2.6\%)} $  \\
\hline
LHC 14 TeV & $953.6$ & $ ^{+22.7 (2.4\%)}_{-33.9 (3.6\%)} $ & $ ^{+16.2 (1.7\%)}_{-17.8 (1.9\%)} $   \\
\hline
\end{tabular}
\caption{\small Our best NNLO+NNLL theoretical predictions for various colliders and c.m. energies.}
\label{tab:results}
\end{center}
\end{table}
Our best predictions are at NNLO and include soft gluon resummation at NNLL \cite{Beneke:2009rj,Cacciari:2011hy}. 

In this letter we take $A=0$ as a default value for the constant $A$ introduced in Ref.~\cite{Bonciani:1998vc}. The reason for switching to a new default value for $A$ (compared to $A=2$ in \cite{Cacciari:2011hy,Baernreuther:2012ws,Czakon:2012zr,Czakon:2012pz}) is that this constant is consistently defined only through NLO. Nonetheless it contributes at NNLO too, and a consistent NNLO treatment would require the analysis of Ref.~\cite{Bonciani:1998vc} to be extended to NNLO, which is now possible with the help of the results derived in this letter as well as Ref.~\cite{Baernreuther:2012ws}. Given the numerical effect is small (a $0.7\%$ shift at LHC 8 TeV and a $0.4\%$ shift at the Tevatron), in this work we take $A=0$.

As can be concluded from table~\ref{tab:results} the precision of the theoretical prediction at full NNLO+NNLL is very high. At the Tevatron, the scale uncertainty is as low as 2.2\% and just slightly larger, about 3\%, at the LHC. The inclusion of the NNLO correction to the $gg$-initiated reaction increases the Tevatron prediction of Ref.~\cite{Baernreuther:2012ws} by about 1.4\%, which agrees well with what was anticipated in that reference. 

\begin{table}[h]
\begin{center}
\begin{tabular}{|c|c|c|c|}
\hline
Collider & $\sigma_{\rm tot}$ [pb]  & scales [pb] & pdf [pb] \\
\hline
Tevatron & $7.009$ & $ ^{+0.259 (3.7\%)}_{-0.374 (5.3\%)} $ & $ ^{+0.169 (2.4\%)}_{-0.121 (1.7\%)} $   \\
\hline
LHC 7 TeV & $167.0$ & $ ^{+6.7 (4.0\%)}_{-10.7 (6.4\%)} $ & $ ^{+4.6 (2.8\%)}_{-4.7 (2.8\%)} $  \\
\hline
LHC 8 TeV & $239.1$ & $ ^{+9.2 (3.9\%)}_{-14.8 (6.2\%)} $ & $ ^{+6.1 (2.5\%)}_{-6.2 (2.6\%)} $  \\
\hline
LHC 14 TeV & $933.0$ & $ ^{+31.8 (3.4\%)}_{-51.0 (5.5\%)} $ & $ ^{+16.1 (1.7\%)}_{-17.6 (1.9\%)} $   \\
\hline
\end{tabular}
\caption{\small Pure NNLO theoretical predictions for various colliders and c.m. energies.}
\label{tab:results-fo}
\end{center}
\end{table}
To assess the numerical impact from soft gluon resummation, in table~\ref{tab:results-fo} we present results analogous to the ones in table~\ref{tab:results} but without soft gluon resummation, i.e. at pure NNLO. Comparing the results in the two tables we conclude that the effect of the resummation is a $(2.2,\,2.9,\,2.7,\,2.2)\%$ increase in central values and $(2.4,\,2.2,\,2.1,\,1.5)\%$ decrease in scale dependence for, respectively, (Tevatron, LHC7, LHC8, LHC14). 
\begin{figure}[b]
\centering
\hspace{0mm} 
\includegraphics[width=0.49\textwidth]{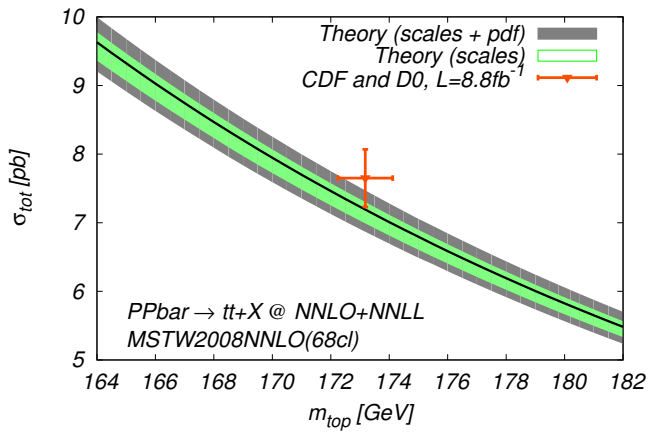} 
\caption{Theoretical prediction for the Tevatron as a function of the top quark mass, compared to the latest combination of  Tevatron measurements.}
\label{fig:tev-abs}
\end{figure}

Next we compare our predictions with the most precise experimental data available from the Tevatron and LHC. 

The comparison with the latest Tevatron combination \cite{tev-sigma_exp} is shown in fig.~\ref{fig:tev-abs}.
The measured value $\sigma_{\rm tot}=7.65\pm 0.42$ pb is given, without conversion, at the best top mass measurement \cite{Aaltonen:2012ra} $m=173.18\pm 0.94$ GeV.
From this comparison we conclude that theory and experiment are in good agreement at this very high level of precision. 

\begin{figure}[t]
\centering
\hspace{0mm} 
\includegraphics[width=0.49\textwidth]{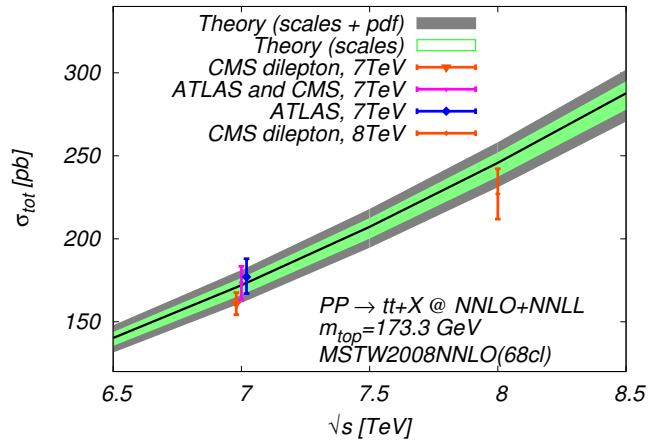} 
\caption{Theoretical prediction for the LHC as a function of the collider c.m. energy, compared to available measurement from ATLAS and/or CMS at 7 and 8 TeV.}
\label{fig:lhc-abs}
\end{figure}

In fig.~\ref{fig:lhc-abs} we show the theoretical prediction for the $t\bar t$ total cross-section at the LHC as a function of the c.m. energy. We compare with the most precise available data from ATLAS at 7 TeV~\cite{Atlas7TeV}, CMS at 7~\cite{Chatrchyan:2012bra} and 8 TeV~\cite{CMS8TeV} as well as the ATLAS and CMS combination at 7 TeV~\cite{Atlas_and_CMS-7TeV}. We observe a good agreement between theory and data. Where conversion is provided \cite{Chatrchyan:2012bra}, the measurements have been converted to $m=173.3$ GeV. 

Finally, we make available simplified fits for the top mass dependence of the NNLO+NNLL cross-section, including its scale and pdf uncertainties: 
\begin{eqnarray}
\label{eq:parametrisation}
\sigma(m) &=& \sigma(m_{ref})\left(\frac{m_{ref}}{m}\right)^4 \\
&& \times \left(1 + a_1\frac{m-m_{ref}}{m_{ref}} + a_2\left(\frac{m-m_{ref}}{m_{ref}}\right)^2\right) \, . \nonumber
\end{eqnarray}
The coefficient $a_{1,2}$ can be found in table~\ref{table:params}.
\begin{table}[t]
\begin{center}
\begin{tabular}{|c|l|c|c|c|}
\hline
\multicolumn{2}{|c|}{${m_{ref}} = 173.3$ GeV} & $\sigma(m_{ref})$ [pb] & $a_1$  & $a_2$\\
\hline
& Central & 7.1642 & $-$1.46191 & 0.945791   \\ 
& Scales $+$ & 7.27388 & $-$1.46574 & 0.957037 \\ 
Tevatron
& Scales $-$ & 6.96423 & $-$1.4528 & 0.921248 \\ 
& PDFs $+$ & 7.33358 & $-$1.4439 & 0.930127  \\ 
& PDFs $-$ & 7.04268 & $-$1.4702  & 0.936027  \\ 
\hline
& Central & 172.025 & $-$1.24243 & 0.890776  \\ 
& Scales $+$ & 176.474 & $-$1.24799 & 0.903768 \\ 
LHC 7 TeV 
& Scales $-$ & 166.193 & $-$1.22516  & 0.858273 \\ 
& PDFs $+$ & 176.732 & $-$1.22501 & 0.861216  \\ 
& PDFs $-$ & 167.227 & $-$1.2586 & 0.918304 \\ 
\hline
& Central & 245.794 & $-$1.1125 & 0.70778  \\
& Scales $+$ & 252.034 & $-$1.11826 & 0.719951 \\
LHC 8 TeV 
& Scales $-$ & 237.375 & $-$1.09562 & 0.677798 \\
& PDFs $+$ & 251.968 & $-$1.09584 & 0.682769 \\
& PDFs $-$ & 239.441 & $-$1.12779 & 0.731019 \\
\hline
\end{tabular}
\caption{\label{table:params} Parameters of the fit (\ref{eq:parametrisation}) to our best NNLO+NNLL prediction, including scale and pdf uncertainty, for the top pair cross section at the Tevatron and LHC at 7 and 8 TeV. The fits are accurate to within few permil in the 130-210 GeV mass range.}
\end{center}
\end{table}

\section{Conclusions and Outlook}

In this work we compute the NNLO corrections to $gg\to t\bar t+X$. With this last missing reaction included, the total inclusive top pair production cross-section at hadron colliders is now known exactly through NNLO in QCD. We also derive estimates for the two-loop hard matching coefficients which allows NNLL soft-gluon resummation matched consistently to NNLO. All results are implemented in the program {\tt Top++}\,(v2.0)~\cite{Czakon:2011xx}.

The theoretical precision achieved in this observable is very high. To illustrate this, we compare to a NLO level calculation. At LHC 8 TeV we observe a decrease in scale dependence by a factor of $(4.3,\,4.2,\,3.0)$ when compared to, respectively, (NLO, NLO+LL, NLO+NLL). The corresponding numbers for the Tevatron are (3.9,\,4.1,\,2.0). 

The predicted $t\bar t$ cross-section agrees well with all available measurements. We are confident that its very high precision will enable a new generation of precision collider applications to, among others, parton distributions and searches for new physics.

\begin{acknowledgments}
We thank S.~Dittmaier for kindly providing us with his code for the evaluation of the one-loop virtual corrections in $gg\to t\bar t g$ \cite{Dittmaier:2007wz}. The work of M.C. and P.F. was supported by the DFG Sonderforschungsbereich/Transregio 9 ÒComputergest\"utzte Theoretische TeilchenphysikÓ. M.C. was also supported by the Heisenberg and by the Gottfried Wilhelm Leibniz programmes of the Deutsche Forschungsgemeinschaft. The work of A.M. is supported by ERC grant 291377 ``LHCtheory: Theoretical predictions and analyses of LHC physics: advancing the precision frontier".
\end{acknowledgments}

\end{document}